\documentclass{aa}
\usepackage{graphics}
\usepackage{psfig}

\begin{document}

   \thesaurus{02.08.1; 	 
	      11.03.1;	 
	      12.03.4;   
	      12.04.1}	 

   \title{On the Matter Distribution of Galaxy Cluster 
          Containing a Compact Core}

   \author{Tzihong Chiueh\inst{1,3} and Xiang-Ping Wu\inst{2}   
          }

   \offprints{T. Chiueh}
   \mail{chiuehth@phys.ntu.edu.tw}

   \institute{$^1$Department of Physics, National Taiwan University,
       Taipei, Taiwan\\
     $^2$Beijing Astronomical Observatory and 
	 National Astronomical Observatories, 
        Chinese Academy of Sciences, Beijing, China\\
      $^3$ Institute of Astronomy and Astrophysics, Academia Sinica, Taipei,
       Taiwan\\ 
    }

   \date{Received June 1999; accepted November 1999}

   \titlerunning{Matter Distribution of Galaxy Clusters}

   \maketitle
   \begin{abstract}

Increasing evidences provided primarily by the cluster lensing and
numerical simulations of cluster formation
indicate that galaxy clusters may contain compact cores that are
substantially smaller than the cores revealed by the X-ray observations 
of hot intracluster gas. In this paper we present a model that describes 
how two distinct cores can grow simultaneously as a result of infall 
from the background dark matter.
This model needs a pre-existing localized large fluctuation,
which can be the non-Gaussian density peak of the primordial
fluctuations, to seed the individual growing cluster.
On the other hand, numerous
recent observations also show that no strong evolution is detected for
galaxy clusters within the redshift up to $z\sim 0.8$.  We
therefore present a comparison of observations with the saturated cluster
configuration resulting from our infall model.
In the saturated state, the predicted compact core mass is about 
few$\times 10^{11} M_\odot$ and the core size about $1$ kpc for a
cluster mass $\sim$ several $\times 10^{15} M_\odot$ within a radius of $3$
Mpc.  We have successfully reproduced the dark matter
distribution revealed by the gravitational lensing, and the observed radial
distributions of cluster galaxies, intracluster gas (i.e., the conventional
$\beta$ model) and baryon fractions in a consistent way.  
This model, when combined
with the observed X-ray surface brightness profiles of clusters, 
predicts that the overall temperature of intracluster gas has a tendency of
radial decline with a mean polytropic index $\gamma\approx1.2$.
Finally, the so-called $\beta$ discrepancy finds a natural
explanation for such a polytropic gas in our model.


   \end{abstract}

%

\section{Introduction}

The existence of a compact and massive dark-matter core in the galaxy cluster,
which is much smaller than the X-ray gas core of the empirical $\beta$ model,
has recently acquired substantial supports from various evidences.
It has been recognized since the early study of  
the very thin arclike images of distant galaxies
created by the potentials of the foreground lensing clusters 
that the dark matter profiles of the lensing clusters should be sharply peaked
towards cluster centers (e.g. Hammer 1991).
In order to account for the total number
of observed giant luminous arcs, the core size of dark matter
distribution modeled by a softened isothermal sphere should be at least
$\sim10$ times smaller than the one for luminous matter profile
(Wu \& Hammer 1993). Grossman \& Saha (1994) strengthened this
argument by explicitly setting an unprecedentedly small upper limit on
cluster core radius, $r_{dark,c}<11$ kpc, for a softened isothermal sphere.
Many recent work on modeling of arcs/arclets have essentially confirmed 
these findings (e.g. Giraud 1999; William, Navarro \& Bartelmann 1999). 
Even for the X-ray measurements of hot intracluster gas, 
the dynamical analysis also arrives at a similar
conclusion that the binding matter distribution of clusters should have a core
radius of $r_{dark,c}<100$ kpc (Durret et al. 1994).
Optically, the omni-presence of
cD galaxies in the cores of many rich clusters may indicate unusual
accumulation of central masses.
Finally, a singular mass profile of $\rho\propto r^{-1}$
at cluster center is predicted by
the numerical studies of cluster formation in the framework of
the standard CDM model as well as other cosmological models
(e.g. Navarro et al. 1995; Cole \& Lacey 1996; Navarro, Frenk \& White 1997; 
Fukushige $\&$ Makino 1997; Eke, Navarro \& Frenk 1998).
Such a singular mass profile also houses a weak compact core.

That is, while the cluster X-ray gas often shows a core of typical size
about $100 - 250$ kpc, the dark matter core can instead be substantially
smaller than $100$ kpc.  The co-existence of two distinct mass cores in a
single cluster appears to have challenged our current
understanding of cluster formation.
Although a great deal of theoretical efforts have in the past been put forth
attempting to grasp an understanding of the cluster physics, few have
considered the possibility of a massive compact dark matter core in the galaxy
cluster.  Past analytic investigations for the gravitational infall of matter
into galaxy clusters have been pursued mainly toward the following two issues:
the cosmic evolution of cluster properties such as the total mass,
number counts, gas luminosity and temperature (e.g.
Gunn \& Gott 1972; Press \& Schechter 1974; Kaiser 1986; 1991),
as well as the gas and dark matter density profiles inside clusters
(e.g. Fillmore \& Goldreich 1984; Bertschinger 1985; Teyssier,
Chi\`eze, \& Alimi 1997).  Regardless of their oversimplification,
these analytic treatments based on self-similar models indeed
provide the gross properties of clusters that were shown to be in gross
consistency with optical/X-ray observations (Kaiser 1991).
Moreover, these analytical studies can also 
reveal the underlying cluster evolution mechanisms, in a complementary manner
to those shown by the particle and hydrodynamic simulations.  In spite of the
past success, previous studies commonly obtained (or assumed) a single
core within the cluster.

In the self-similar regime, clusters are, however, shown to undergo very
rapid evolution
(Gunn \& Gott 1972; Kaiser 1986).  When these results are combined with the
recent observations made primarily at X-ray wavebands, which demonstrate
that no evolution has been detected within the redshift at least out to
$z\approx0.8$ among the cluster number counts, X-ray luminosity, temperature
function as well as typical scale lengths (e.g. Fan, Bahcall \& Cen 1997; 
Rosati et al. 1998;
Vikhlinin et al. 1998; Wu, Xue \& Fang 1999  and references therein), 
one is led to a conclusion
that the evolution of these clusters must have ended at an early epoch.

In this paper, we will concentrate on the dark matter and the
galaxy/gas distributions of clusters revealed mainly by the saturated state
of a new class of self-similar infall model,
which describes the simultaneous build-up of two distinct
cores in a single cluster.
Our starting point differs from the past
theoretical models in the following way: (a) Unlike the previous work which
chose a power-law initial perturbation in the mass (e.g. Fillmore \& Goldreich
1984; Teyssier et al. 1997), we seek a solution to the
infall problem of dark matter particles by assuming
an initial compact seed density fluctuation at the site of growing cluster.
This aspect is in line with the works of Gott (1975) and Bertschinger (1985),
and the large-amplitude seed fluctuation may arise from a
non-Gaussian peak in the primordial density fluctuations.  
(b) Instead of assuming the collisionless particles to move with 
radial orbits (Merritt, Tremaine \& Johnstone 1989), 
which are known to be unstable (Binney \& Tremaine 1987), we
consider the other extreme situation where the particle orbits inside the
cluster are so chaotic that the velocity dispersion is essentially isotropic.  

After a substantial growth of both cores,
they can rapidly reach a saturated state, in which the mass ratio of
the two cores
remains approximately the same as that during the infall
phase. (This is because both
cores are gravitationally bound, and they have been grossly
virialized regardless of whether there exists an infall.)  Our
attempt at comparing the theoretical predictions with observational results
will therefore be made at this late stage when the matter
infall has ended, and a hydrostatic equilibrium has already been
established between the dark-matter gravitational potential and
the cluster galaxies and intracluster gas.

We begin in section 2 by presenting a specific model describing the
self-similar evolution of spherical collapse of collisionless cold
dark matter in an open and a closed universe.  
(The former also represents the situation where the infall began
only recently and is still ongoing in a flat universe.)
We then derive
the density profiles of
dark matter in the infall phase and in the saturated phase, respectively.
In section 3 the predicted dark matter distribution and
the galaxy/gas density profiles are compared with those revealed by
the gravitational lensing and optical/X-ray observations.  Finally, the
discussion and summary are given in section 4.
Beginning from section 3,
where comparisons of the theoretical predictions with the observations are
made, we work mainly with the saturated state (except in section 3.1), 
and the Hubble constant is chosen to be $H_0= 50$ km s$^{-1}$ Mpc$^{-1}$ for
convenience.  Throughout this work, we set the cosmological constant equal
to zero.

\section{Self-similar solutions for the collapse of cold dark matter}

\subsection{Physical Picture and Mathematical Formulation}

Assuming the linear growth of an overdense spherical top-hat perturbation,
Gott (1975) and Bertschinger (1985) have derived
that a bound object can grow, as a result of matter infall from the
pressureless Hubble flow, with the time dependence:
$M\propto t^{2/3}$ and $r_m\propto t^{8/9}$,
where $M$ is the total mass within the radius $r_m$ of the growing bound
object and $t$ is the evolution time.  Such a scaling relation depends 
only weakly
on the decelerating parameter $q_0$ (for $0.02 < q_0 < 2$), and these
relations become exact when $q_0=1/2$.
A slightly different scaling relation has also been predicted by
Press and Schechter
(1974) in their quest for the galaxy mass function.  They found that
$M\propto t^{4/3}$ and $r_m\propto t^{10/9}$.  A
feature common to both scaling relations is that the characteristic length
scale $r_m$ is approximately a linear function of
time.  This behavior will be used to fix the temporal scaling of velocity
as $t^0$ in our self-similar model.

On the other hand,  the crossing of mass
shells near the core can trigger particle orbit instabilities that yield
violent relaxation (Lynden-Bell 1967).   The particle orbits may be further
randomized by the interactions with 
the crowded discrete galaxies in the cluster,
making the dark matter particles further mixed.  
We therefore approximate the dark matter within the cluster
to have an isotropic and also spatially uniform velocity dispersion
$\sigma_d$.  Together with the temporal scaling discussed in the last
paragraph, one may consider $\sigma_d$ 
to be a constant in space and in time within the cluster.  (For readers
who have a different view of how the dark matter particles should
behave in the cluster,
they may simply regard the constant $\sigma_d$ as a working hypothesis 
adopted to fix the self-similar scaling relations.)
Outside the cluster, the cold dark matter detached from the
expanding background Hubble flow falls radially inwards.
The boundary of the hot cluster and cold background is defined
by a heat conduction front, across which both mass flux and momentum flux are
continuous.  

Thus, our picture of the self-similar matter infall can be summarized as
follows.  As the cold dark
matter falls towards the massive compact core, a volume of finite radius
surrounding the compact core then undergoes violent relaxation
and forms an isothermal halo, which gives rise
to the cluster potential.  A small fraction of the infall particles can
however sink into the compact core and therefore the peak of the central
density can grow steadily (Gott 1975).  But the majority of infall
particles are deflected back to the halo.  The deflected
particles find the halo to become more massive than they previously experienced
due to the continuing infall, and hence the particles can
travel only up to a maximum radius, which defines the 
location of the conduction front, i.e., the boundary of the hot cluster.  As
these processes continue, both compact core and halo can grow
self-consistently.

To put this picture into a quantitative perspective, we adopt a
spherically symmetric infall model. 
Since the particle orbits are assumed to
be isotropic in the cluster bulk, the anisotropic stress force is unimportant
and it suffices to consider only the radial pressure force in the momentum
balance.  The unique self-similar scaling with a constant $\sigma_d$ requires
the mass density of dark matter particles $\rho(r,t)$, the flow speed $u(r,t)$ 
and the total mass $M(r,t)$ enclosed within radius $r$ to be scaled as:
\begin{equation}
\rho(r,t)=\frac{\alpha(x)}{4\pi G\tau^2}, \;\;
u(r,t)=\sigma_d v(x),\;\;
M(r,t)=\frac{\sigma_d^3\tau}{G}m(x),
\end{equation}
where $\tau \equiv t$ and $x$ ($\equiv r/\sigma_d \tau$)
is the distance in a comoving frame.  
The continuity equation in the comoving frame then becomes 
\begin{equation}
-x^2(2\alpha +x\frac{d\alpha}{d x})+\frac{d}{d x}(x^2 v\alpha)=0.
\end{equation}
This equation can be rewritten as $d(x^2(x-v)\alpha)/dx=x^2\alpha$.
Comparing this equation 
with the relation between $\alpha$ and
$m$,  i.e., $dm/dx=x^2\alpha$,  we find that the mass $m$ can be expressed 
as $m(x)=x^2(x-v)\alpha$.   This expression of $m(x)$ is substituted
into the gravitational force to eliminate $m(x)$ in the momentum equation.  
Re-arrangement of the momentum equation 
\begin{equation}
(v-x)\frac{d v}{d x}=-\frac{1}{\alpha}\frac{d\alpha}{d x}-(x-v)\alpha,
\end{equation}
and eq.(2) yields
\begin{eqnarray}
\left[(x-v)^2-1\right]\frac{dv}{dx}=
	   \left[\alpha(x-v)-\frac{2}{x}\right](x-v),\\
\left[(x-v)^2-1\right]\frac{1}{\alpha}\frac{d\alpha}{dx}=
		      \left[\alpha-\frac{2}{x}(x-v)\right](x-v).
\end{eqnarray}

Note that eqs.(4) and (5) have been derived in the past for describing the
inside-out collapse of a singular isothermal molecular cloud (Shu 1977).
However, our boundary conditions are different from those for the
molecular cloud, in that the hot cluster is fed from outside by the expanding
cold dark matter, and inside, the collisionless matter lands onto
the compact core at a vanishing speed so that the particles can be captured.
By contrast, the collapsing
molecular cloud is fed from outside by a static isothermal gas sphere;
inside, the collapsing gas lands on the protostar
violently with a Mach number much greater than unity, and it relies eventually
on the radiative cooling for the infall matter to be captured by the prostar.

Eqs.(4) and (5) can be simplified when the dark matter
is cold, by setting the pressure term on the right of eq.(3) to zero.  It
yields 
\begin{eqnarray}
\frac{dv}{dx}=\alpha\\
(x-v)\frac{1}{\alpha}\frac{d\alpha}{dx}=\alpha-\frac{2}{x}(x-v).
\end{eqnarray}
These equations are valid for the background dark matter particles
before falling within the cluster.   Note that both eqs.(4) and (5), and
eqs.(6) and (7) allow for the Hubble-flow solution of a flat universe 
where $\alpha=2/3$ and $v=2x/3$.

Finally, the most relevant regime consistent with the current observations
corresponds to a final saturated and static
configuration when the matter infall ended before the present.
That is, most
of the available dark matter particles have collapsed into the
cluster.  Under these circumstances, such a saturated system is
completely determined by the static forces in eq.(3):
\begin{equation}
\frac{1}{\alpha}\frac{d\alpha}{dx}=-\frac{m}{x^2}.
\end{equation}
Here, we have kept the same dimensionless parameterization although
the dark matter density and mass are now time-independent.  The additional
factor $\tau$ in this dimensionless parameterization should be interpreted as
the duration of cluster formation.

\subsection{Infall boundary conditions}

Our inner boundary condition of eqs.(4) and (5)
is so chosen that a growing massive compact core is present.
For the collapsed particles to
settle into the compact core, we demand that the cluster bulk flow
should have a vanishing infall speed at the core.  Quantitative
details as to how the matter falls into the compact core will be further
elaborated below.

Far from the cluster, we demand the pressure-free dark
matter solutions to obey eqs.(6) and (7), and they must either asymptotically
match to a vanishing-density background when $\Omega <<1$,
or asymptotically
match to the Hubble flow ($\alpha=2/3, v=2x/3$) when $\Omega=1$.
These pressure-free, outer infall solutions are then matched to the interior
subsonic cluster solutions of eqs.(4) and (5)
through a co-moving conduction front.  Technically, we must adjust both the
location and strength of the conduction front to determine a unique
solution satisfying both inner and outer boundary conditions.

\subsection{Solution in an open universe ($\Omega <<1$) or the 
ongoing infall case}

The background density of Hubble flow in this regime assumes a value
$\alpha << 2/3$,
and our outer infall solution should asymptotically decrease to match this
low-density background far from the cluster.  Practically, this boundary
condition demands the asymptotic outer infall solution to vanish
at large distances.

We shall stress that the solution to the open-universe case can also represent
the situation where the self-similar infall began only recently in a flat
universe, in which case $\tau$ is redefined as $t-t_1$, with $t_1$ being the
starting time of infall and $\tau$ much less than the cosmic age $t$.   As
will be shown in the following analysis, the conduction front has a density
$\rho\sim 1/4\pi G\tau^2$, much greater than the background density,
e.g., the critical density $\rho_c (= 1/6\pi Gt^2)$.  Therefore the outer cold
infall solution should asymptotically match the vanishingly small background
density at large distances.  Such a boundary condition is exactly the same as
the open-universe case.  We shall briefly return to this case in section
3.2 for a comparison with observations.

The outer pressure-free solution is constructed from a remote location far
from the cluster by integrating eqs.(6) and (7) inward.
To ensure that the solution beyond the starting distance of
integration indeed asymptotically matches onto the background flow, we
perform an asymptotic analysis for the large-distance solution.
The large-distance solution of eqs.(6) and (7) is $\alpha=c_0/x^2$
and $v=-c_0/x$.    That is, the density
$\alpha$ and velocity $v$ must be related in this specific
way to warrant that the density asymptotically decreases to a sufficiently
low value and matches onto the low background density.  
However, since eqs.(6) and (7) are
actually nonlinear equations, the amplitude
$c_0$ must also be constrained in such a way that the outer infall solution
be matched to the interior hot cluster solution through a
conduction front.  In this regard, the amplitude $c_0$ can be taken as a
nonlinear eigenvalue imposed at the outer boundary.
Moreover, the interior cluster solution is demanded to possess a
diminishing velocity when the flow enters the compact core,
so that the infall matter can be captured by the growing compact core.
This requires a unique location for the conduction front as the
second nonlinear eigenvalue.

To meet these non-trivial conditions, we find it useful technically 
to search for an approximate solution of eqs.(4) and (5), which makes a
smooth transition from a subsonic infall in the bulk to a supersonic infall
into the core.   This is because the trans-sonic solution is at the
bifurcation boundary in the solution space (Shu 1977), and our desired
solution is in the neighborhood of this boundary.  The desired
solution can then be constructed relatively easily after the approximate
solution is found.  Conditions for a flow to smoothly pass
through the sonic speed are $\alpha_s=2/x_s$ and $v_s=x_s-1$, where the
subscript $s$ denotes the sonic point which is a critical point of eqs.(4) and
(5). 

By adjusting both the amplitude $c_0$ of the asymptotic background
solution and the location $x_s$ of the sonic critical point,
one may uniquely determine the location of
and density jump at the conduction front, across which the mass flux and
momentum flux in the co-moving frame are continuous.
Our numerical search shows that the approximate solution has 
$c_0\approx 1.22$
and $x_s\approx 10^{-3}$, and the location of the conduction front at
$x_{cond}\approx 2/3$.
Upstream of the conduction front, the dark matter density
$\alpha \approx 1.6$ and infall speed $v\approx -1.45$,
whereas downstream of the front, $\alpha\approx 3.5$ and $v\approx -0.32$.
This conduction front has a moderate compression ratio $\approx 2.2$.

An eigenvalue search in the neighborhood of this trans-sonic solution 
permits us to obtain the desired solution easily, which turns out
to differ from the trans-sonic solution only near the compact
core.   As previously stated, the transonic solution is the bifurcating solution,
and it has an infall speed increasing inward moderately.  On one
side of this bifurcation boundary, the infall speed increases drastically
inward, but on the other side the infall speed 
turns around to become vanishingly small toward the core.  (See Fig. 1.)  
(These situations are not unlike the behaviors of the scaling factor $a(t)$ in 
a flat
universe, an open universe and a closed universe.)  Clearly, the latter is what
we look for.  As the desired solution and trans-sonic solution
are almost identical in most part of the cluster, their conditions at
the conduction front are practically the same.

A caveat for our results is that the collisionless conduction
front is likely to have a transition layer of finite thickness. Hence the
results given above
can only be regarded as a sophisticated estimate for modeling the boundary of
the cluster, and the discontinuity obtained here should not be taken too
literally.

The derived density profile of dark matter flow can be fit nicely by an
analytical form:
\begin{equation}
\alpha(x)=\left\{
\begin{array}{ll}
\frac{1.15}{(0.13^2+x^2)^{3/2}}\;e^{0.0015/x};
				    & 10^{-3}<x<x_{cond};\\
				    &		\\
\frac{1.13}{0.52^2+x^2} ; &10> x>x_{cond}.
\end{array}\right.
\end{equation}

Two separate components for the interior cluster solution are identified.  The
first component $(\propto e^{0.0015/x})$ gives rise to a peaked density
profile, in response to the gravity of massive compact core; the second
component represents the usual Bonner-Ebert state of isothermal sphere, which
has a secondary core of size $x_c\approx 0.13$, and the density outside the
outer core decreases outward first as $x^{-3}$ and subsequently as $x^{-2}$.
The conduction front is located roughly at the transition between the $x^{-3}$
and $x^{-2}$ profiles. The matter located in the region from $x\sim 0.02$ to
$x\sim 0.4$ can be well approximated by a ``static
sphere'', where the flow speed is roughly zero.  Near the conduction front the
infall speed sharply rises, and the finite flow speed immediately downstream of
the conduction front serves only to flatten the density profile slightly.
At about $x=0.02$, the mass within this sphere is twice
as much as that of the compact core; that is, the gravity arising from
the massive compact core begins to be overwhelmed by the self-gravity of the
matter outside the compact core.
Figs.1(a), (b) and (c) depict
the normalized density, velocity and mass profiles, respectively.   Note that
most part of the hot cluster is almost static, except near the conduction front
and immediately outside the compact core.

    \begin{figure*}[t]
      \psfig{figure=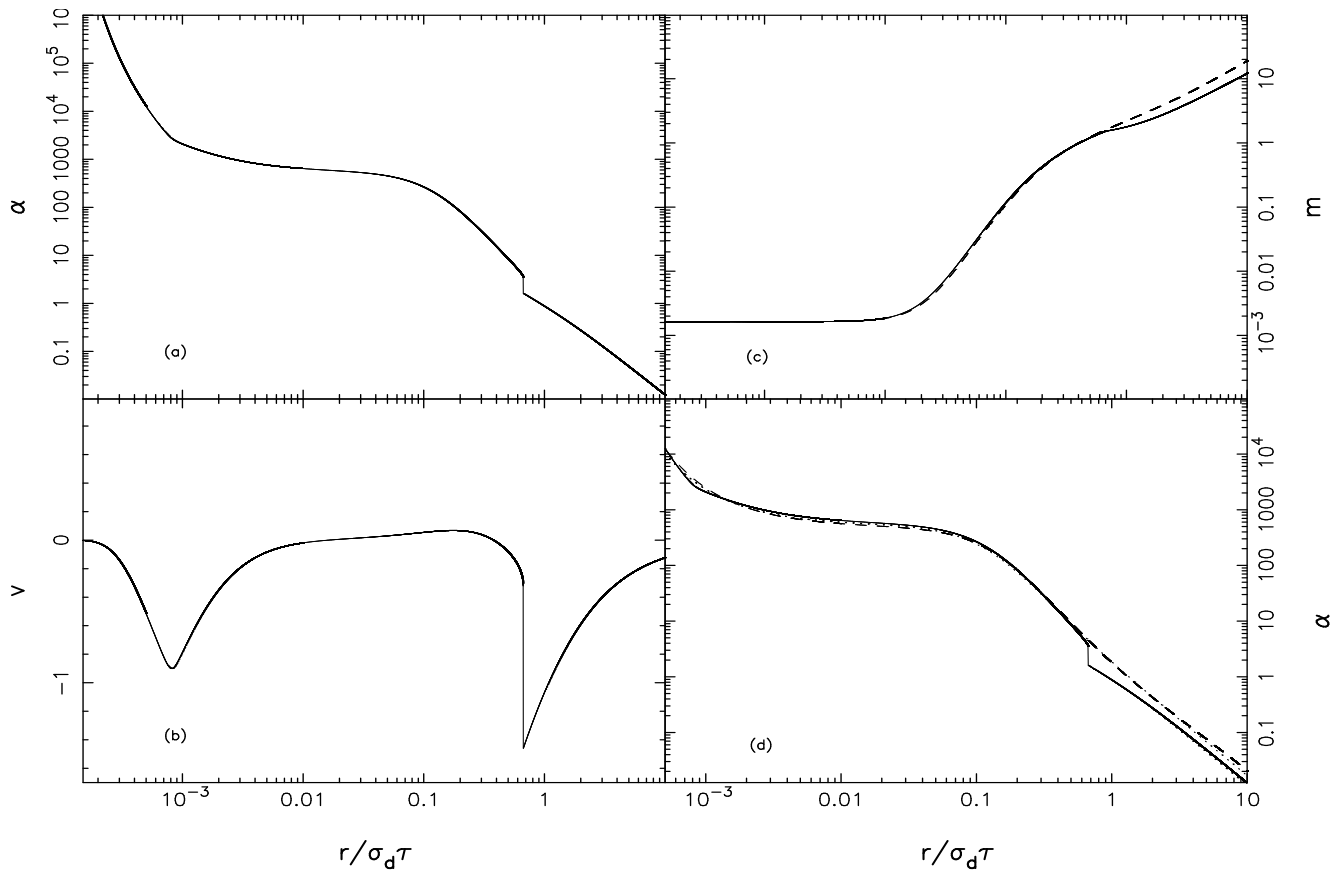,height=120mm,width=\textwidth,bbllx=50pt,bblly=50pt,bburx=580pt,bbury=756pt,clip=,angle=270}
	\caption{(a)Dimensionless density profile, (b)flow speed and (c)
	cluster mass obtained from the recent-infall, self-similar
	solutions under the initial and boundary conditions of a mass
	concentration $m_0=1.6\times10^{-3}$ within $x=r/\sigma_d\tau=0.0005$.
	Also shown are the density (d) and mass (c) variations
	for a saturated cluster (dashed lines) expected under the same 
	inner boundary condition. The dotted lines in (d) are the analytic 
	fits of the density profiles by eqs.(9) and (10), respectively.
}
      \label{fig.1}
   \end{figure*}

\subsection{Infall into the compact core}

The compact core is a virialized bound object, where the density rises
inwards more steeply than $r^{-3}$.  As shown in Fig.(1) when the dark
matter particles
fall within $x=0.02$, they experience solely the core gravity and are
accelerated inward
to reach a maximum speed at $x\sim 8\times 10^{-4}$.  Inside $x\sim 8\times
10^{-4}$, the infall
suddenly gets retarded by the core pressure and its speed decreases
from this point on and becomes vanishingly small at the core center.

To pin
down where the true core boundary is, we examine the binding energy.
Inside the core, the flow must satisfy the binding condition,
$|P.E.|\geq 2|K.E.|$, so that no particle may escape outside the radius $x$
once enters.
The specific potential energy $P.E.$ at the radius $x$ is $-m/x$, and the
specific kinetic energy $K.E.$ includes the specific thermal energy $3/2$ and
the comoving flow energy $(v-x)^2/2$.
Our solution shows that only inside $x=5\times 10^{-4}$ can the binding
condition be satisfied.  The size of the compact core is thus set to be
$x_b=5\times 10^{-4}$.

\subsection{Solution of early infall in a flat universe}

The asymptotic solution far away from the cluster behaves as
$v=(2x/3)-c_1/x$ and $\alpha=(2/3)+c_1/x^2$, when the deviation from
the Hubble flow is small.  We may employ the technique used in
the previous case to obtain the solution in this regime.  Surprisingly,
the trans-sonic solution yields a conduction front at almost the same location
and with a comparable jump as those of the previous case.  The nonlinear
eigenvalue $x_s$ for the trans-sonic solution is also the same.
In addition, the nonlinear eigenvalue $c_1$ also assumes a
value equal to $c_1=1.22$.   As a result, we find that the desired solution
interior of the cluster is nearly the same as that of the previous case.
The outer cold infall solution is, however, different from that in eq.(9);
the following analytical expression is found to fit the outer cold infall
solution:
\begin{equation}
\alpha(x)=\frac{0.91}{x^{3/2}}+\frac{2}{3};
\ \  \ \ 10>x>x_{cond}.
\end{equation}

\subsection{Saturated phase}

Taking the typical dark-matter velocity dispersion, 
$\sigma_d\approx$ 1000 km s$^{-1}$, and the typical size of
an Abell cluster, $3$ Mpc, it follows that
the typical time scale for cluster growth is on the order of $10^9$ years,
a short time compared with the age of the universe.  Moreover,
no strong cosmic evolution for clusters since $z\sim 0.8$ has been detected
by recent X-ray observations (Rosati, et al.
1988; Vikhlinin et al. 1998).
It is, therefore, reasonable to assume that most clusters have reached
saturated states by $z\approx 0.8$.  The comparison of observations with
the present model is to be made primarily for
the saturated state, where the clusters are virialized.

Because the central compact
core is strongly gravitationally bound, the core
should remain intact in the saturated phase, and the core size and mass
are frozen to their values immediately before the infall stopped.
Note that the extra factor $\tau$ that makes eq.(7) dimensionless can be 
interpreted as the cosmic age at which the infall ended for an early 
formed cluster.  Therefore, we may conveniently choose
the same values of compact core mass ($m=1.6\times
10^{-3}$) and size ($x_b=5\times 10^{-4}$) as those of the infall phase
for the static cluster solution.

The integration
of eqs.(2) and (7) can be carried out from the compact core outward
to a sufficiently large distance to obtain the static cluster solution.
The solution construction is straightforward once the core size and mass are
specified.
The resulting density profile of dark matter is plotted in Fig.1(d) and
can also be fit by an analytical form:
\begin{equation}
\alpha(x)=1.68\;e^{0.0015/x}\;\frac{\sqrt{0.53^2+x^2}}
				{(0.12^2+x^2)^{3/2}}.
\end{equation}
The density near the core also behaves as $\propto e^{0.0015/x}$, and
again has a secondary outer core $x_c=0.12$, beyond which the
dark matter decreases outward also from a $\propto x^{-3}$ profile to 
$\propto x^{-2}$ profile at
large distances.  The corresponding profiles are depicted in Fig.1(c \& d).

In sum, for the solutions of both ongoing-infall (open universe) and
early-infall (flat universe) cases, the interior hot clusters are found to be
almost identical.  This result to some degree reinforces the assertion of Gott
(1975) that the self-similar evolution of bound objects depends weakly on the
deceleration parameter.  In the saturated phase, we have assumed that the
compact core is frozen to the configuration immediately before the infall
ended, and obtained a saturated cluster configuration that departs only
slightly from those of the infall phase. We also find that the plausible value
for the compact core radius approximately equals $x_b\approx
5\times 10^{-4}$ and the core mass, $1.6\times 10^{-3}$.  Taking a typical
cluster for which $x=1$ corresponds roughly to 1 Mpc and a typical velocity
dispersion $\sim$1000 km s$^{-1}$, we estimate $x_b\sim$ 0.5 kpc and the
compact core mass about $3\times 10^{11} M_\odot$.

\section{Comparison with observations}

\subsection{Rationale and strategy}

In this section, the predicted dark-matter distribution in the galaxy cluster 
will be tested against
the observed results of gravitational lensing, galaxy distribution, 
gas distribution and baryon fractions.  The rationale behind such a comparison 
is that if the theoretical
prediction can well approximate the true dark-matter distribution, a single
gravitational potential given by the predicted dark-matter distribution 
ought to be able to produce consistent profiles for these different 
observations.  Our strategy is therefore 
first to use the gravitational lensing
data to calibrate the length scale and depth of the gravitational potential,
which are related to the only two parameters, $\tau$ and $\sigma_d$, in the
theoretical model.  Once the potential is fixed by the lensing data, 
the baryons will be
treated as test particles, which should be so distributed as to be in
dynamical equilibrium with the potential. 

However, the construction of galaxy and gas distribution also requires 
an additional knowledge of their ``equations of state''.  
For galaxies, we adopt the 
average profile of the observed galaxy velocity dispersion and for gases, 
a polytropic equation of state.  Though it can be relatively 
easy to find the best
galaxy and gas parameters for a theoretical model to compare well with
the observations, we would like to go one step beyond, and to constrain the choice of these 
empirical parameters based on a physical principle.  That is, 
we consider the dark matter particles, galaxies and gases to have already been
grossly virialized, and they should
share a common {\it averaged} velocity dispersion since they are all 
within a common gravitational potential.  It is under 
this additional constraint that the comparisons will be made.

\subsection{Dark matter distribution: comparison with the gravitational
lensing results}

Gravitational lensing appears to be a powerful and unique tool at present
for probing the total matter distribution of galaxy clusters.
It provides the gravitating masses of clusters regardless of
the matter compositions and dynamical states. So far,
the strongly and weakly distorted images of background galaxies have
been detected in about 50 clusters at intermediate redshifts ranging
from $z\approx0.1$ to $z\approx0.8$. These clusters constitute a
good sample of matter distributions on scales $r\sim10$ kpc -- $\sim1$ Mpc,
both individually and statistically
for testing various mass density profiles advocated by theoretical and
observational studies.  A direct comparison between
our solution eq.(11) and the gravitational lensing results
is straightforward, provided that the evolutionary time $\tau$ and the
velocity dispersion $\sigma_d$ of dark matter particles are known.
On the other hand, such a comparison will allow us to fix these
two free parameters, if the self-similar solution predicts a consistent
mass profile with what is derived from the gravitational lensing.   
Due to the cosmic variance, these two parameters can only be fixed sensibly by the averaged mass profile of the lensing results. 

Before we fix these parameters by the averaged lensing results, a comparison
with two individual clusters is presented to illustrate how well the theoretical 
profile may agree with the observed profiles individually, and not just statistically.   
Fig.2 shows the projected radial mass distributions of two distant
clusters at $z\approx0.8$, MS1137+66 and RXJ1716+67,
obtained recently by employing the mass reconstruction technique
for the detected weak distortions of distant background galaxies
(Clowe et al. 1998).  Also plotted are the 2-D mass profiles
from eq.(11) for ($\sigma_d$, $\sigma_d \tau$)=(850 km s$^{-1}$, 1.5 Mpc)
and
($\sigma_d$, $\sigma_d\tau$)=(1000 km s$^{-1}$, 0.6 Mpc), respectively.
(From now on, the quantity $\tau$ should, unless otherwise indicated, be
interpreted as the cluster formation time scale.)
It appears that the self-similar solution does yield an acceptable
fit to the results of MS1137+66 and RXJ1716+67.
Moreover, the required values of velocity dispersion,
$\sigma_d\sim1000$ km s$^{-1}$,
for dark matter are also in good agreement
with the observed X-ray luminosity of the two clusters,
$L_x\approx (6-8)\times10^{44}$ erg s$^{-1}$ in the energy band
0.3--3.5 keV. Meanwhile, the scale length $\sigma_d \tau\sim1$ Mpc
indicates a comfortable cluster formation time of $\tau\sim10^{9}$ yrs.

    \begin{figure} 
	\psfig{figure=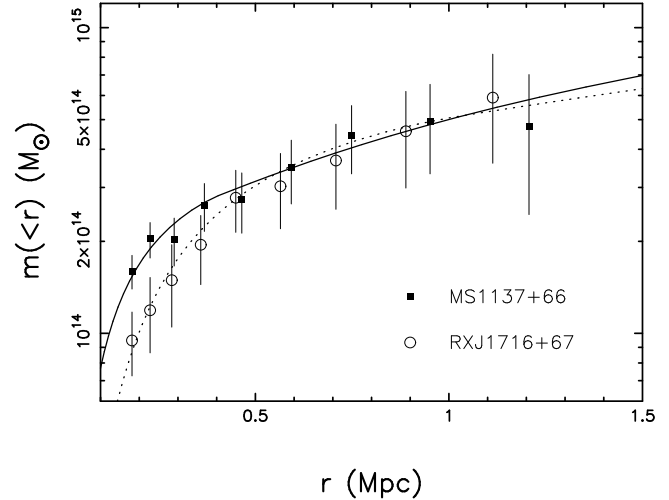,width=88mm,bbllx=86pt,bblly=86pt,bburx=564pt,bbury=642pt,clip=,angle=270}
       \caption{A comparison between the projected cluster masses 
         derived from the weakly distorted images of background galaxies 
         behind clusters MS1137+66 and RXJ1716+67 and those from our 
         saturation model. Data sets of the lensing cluster masses are 
         from Clowe et al. (1988). Our best fitting  yields 
         $(\sigma_d,\sigma_d\tau)=(850$ km s$^{-1}$,1.5 Mpc)
         and $(\sigma_d,\sigma_d\tau)=(1000$ km s$^{-1}$,0.6 Mpc) 
         for MS1137+66 and RXJ1716+67, respectively.
}
      \label{fig.2}
   \end{figure}

This apparent success encourages us to conduct a comparison
between the overall radial mass distribution of clusters
revealed by the strong/weak lensing and that predicted by
our analytical solutions. Note that in most cases,
only the rich and X-ray luminous clusters of galaxies
at intermediate redshifts are capable of gravitationally distorting
the background galaxies.  Namely, the (strong) lensing clusters alone
are a fair sample of massive clusters at $\overline{z}\approx0.3$
in the universe. We display in Fig.3 the mean projected cluster mass
$m_{lens}$ within $r$, derived from a total of 49 arcs/arclets in 38 clusters
and 134 weak lensing measurements made among 24 clusters (Wu et al.
1998 and reference therein), in which we have not included the result of
A2163. While there may exist a discrepancy
between the strong and weak lensing measured cluster masses,
the cluster masses obtained from three different lensing phenomena,
including the deficit of red galaxy population detected behind A1689
due to the lensing magnification (Taylor et al. 1998), are essentially
consistent within $2\sigma$ error bars.
The radial mass variation of the lensing clusters from
our self-similar solution with $\sigma_d=1300$ km s$^{-1}$ and
$\sigma_d\tau=1$ Mpc is also demonstrated in Fig.3. The good
fitness of our prediction to the lensing result is clearly seen.
Since the velocity dispersions of optical galaxies in lensing clusters
are in the range of 600 -- 2000 km $^{-1}$ with an average value of
1200 km s$^{-1}$ (Wu et al. 1998), the required mean velocity dispersion
of dark matter particles $\sigma_d=1300$ km s$^{-1}$ is more than acceptable.
Consequently, the mean formation time of the lensing and massive clusters
turns out to be $\tau=7.5\times10^{8}$ yrs.  This timescale seems
to be relatively short though it is not impossible for the very
rich clusters (Gunn  \& Gott 1972).

    \begin{figure} 
	\psfig{figure=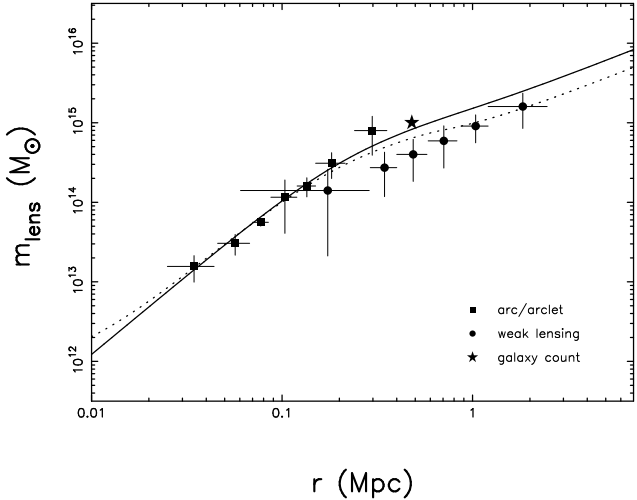,width=88mm,bbllx=86pt,bblly=86pt,bburx=564pt,bbury=642pt,clip=,angle=270}
	\caption{An overall radial variation of the projected cluster 
          masses revealed by the gravitational lensing methods:
          the strongly (arcs/arclets) and weakly distorted images of
          background galaxies (for references see Wu et al. 1998)
          as well as the deficit of red galaxies behind A1689 
          (Taylor et al. 1998). The results predicted by the recent-infall 
          solution (dotted line) and the saturated configuration 
          (solid line) are shown for $\sigma_d=1300$ km s$^{-1}$ 
          and $\sigma_d\tau=1$ Mpc.
}
      \label{fig.3}
   \end{figure}

With the same velocity dispersion and scale length parameters,
our theoretical results produce equally good fits for
the mass profiles given by both the static solution of eq.(11)
and the collapsing solution of eq.(9).	 The latter should now be interpreted
as the solution of a growing cluster which began its formation only recently.
Also shown
in Fig.3 is that the difference between the mass profiles is only
minor, leading to an inability to observationally distinguish the two models
simply from their mass profiles.  To differentiate these two models, it is
helpful to recall some established observational evidences
of galaxy clusters. It was shown more than a decade ago that optical counts of
clusters did not exhibit any evolutionary tendencies for redshift
out to at least $z\approx 0.5$ (Gunn, Hoessel, \& Oke 1986).
The same conclusion holds true for the X-ray selected clusters
since redshift of as high as $z\sim0.8$ (Rosati et al. 1998 and references
therein).  Furthermore, no significant differences in the
dynamical properties between low-redshift and high-redshift clusters,
such as the
X-ray luminosity-temperature relationship, the velocity
dispersion-temperature relationship (Mushotzky \& Scharf 1997; Wu et al. 1999),
and the baryon fraction (see Fig.7), have been detected.
In particular, the distribution of core radii of the intracluster
gas in nearby clusters is identical to that of distant clusters
($z>0.4$) (Vikhlinin et al. 1998).
This last point can be regarded as the most convincing
evidence for the already ``settled'' configuration of cluster matters,
but sharply discords with an evolving self-similar model, which
predicts the core radius to be much smaller in the past (Kaiser 1986).
Suggested by these evidences
can be that significant matter infall processes should have already ended
before
$z\sim 0.4$ so that the global properties of galaxy clusters have remained
approximately unchanged since an even earlier epoch.

\subsection{Galaxy distribution}

The rationale of the following exercise goes as follows. From 
the massive-cluster samples measured by
gravitational lensing, we have statistically determined the
mean velocity dispersion
$\sigma_d (\approx$ 1300 km s$^{-1}$) and
the mean length scale $\sigma_d \tau$ ($\approx 1$ Mpc),
thereby fixed the average mass and length scales of our solutions
for massive clusters.
This predicted average solution hence has {\it no} adjustable parameter.
So, we may employ the Jeans equation to construct an expected
average galaxy distribution and compare it with the
observed average galaxy distribution within the massive clusters.

So far we have not yet included the baryonic matter
(galaxies and gas) in our self-similar solutions
of dark matter particles in clusters.  Although an exact self-similar
solution with a proper combination of the collisionless dark matter
and the collisional gas infall can be constructed,
we would rather take a less
vigorous approach to this problem by treating the baryons as test particles.
For galaxy populations
characterized by their radial velocity dispersion $\sigma_{gal}(r)$ and
number density $n_{gal}(r)$, the Jeans equation under the
light-traces-mass hypothesis reads
\begin{equation}
\frac{1}{n_{gal}}\frac{d(\sigma_{gal}^2 n_{gal})}{dr}
+\frac{2\sigma_{gal}^2}{r}A =
-\frac{GM}{r^2},
\end{equation}
where $M$ is the total cluster mass (we use dark matter instead)
enclosed within
radius $r$, and $A$ denotes the velocity anisotropy parameter.
Since the above equation corresponds to a purely hydrostatic equilibrium,
we employ the static solution, eq.(11), to evaluate $M(r)$.
(The collapsing solution, eq.(9), has also been tested, and
essentially the same $M(r)$ is obtained.)
Once the velocity profile $\sigma_{gal}(r)$
and $A$ are specified, we can easily work out
the galaxy number density profile according to eq.(12).

We compare our prediction with
the average surface number density of galaxies over 14
distant clusters measured by the Canadian Network for Observational
Cosmology (CNOC)
(Carlberg, Yee, \& Ellingson 1997). Actually,
the majority of the 14 clusters are those massive lensing clusters
discussed in the above subsection. We adopt the simple form of
the radial velocity dispersion suggested by CNOC, $\sigma_{gal}^2(r)
=\sigma_{gal}^2(0)/(1+r/r_{200})$ and the assumption for a zero anisotropy
parameter, $A=0$.  Here, $\sigma_{gal}(0)$ is the
central velocity dispersion  and $r_{200}$ denotes
the radius within which the mean cluster mass density is 200 times the critical
mass density of the universe ($\Omega_0=1$).  The mean value of
$r_{200}$ over the 14 CNOC clusters has been found to be
$r_{200}=2.5$ Mpc.

With the predicted gravitational
potential of the underlying dark matter and the given profile of
observed velocity dispersion of galaxies, one can solve for the average
galaxy distribution using the Jeans equation, eq.(12).  In the following
comparison, we will instead take an alternative route by first seeking the best fit parameters and then examine how well
they may agree with those fixed by the lensing results.
Fig.4 shows the CNOC galaxy surface number density
$\Sigma_{gal}$ and the predicted galaxy surface number density from our model
for the best fit $\sigma_d/\sigma_{gal}(0)=0.78$ and $\sigma_d \tau/r_{200}=0.4$.
When combined with
$r_{200}=2.5$ Mpc, this choice of length scale $\sigma_d\tau(=0.4r_{200}=1$ Mpc)
is what has previously been determined by the gravitational lensing.   The good
fit of the predicted length scale with the observed length scale is remarkable.

    \begin{figure} 
	\psfig{figure=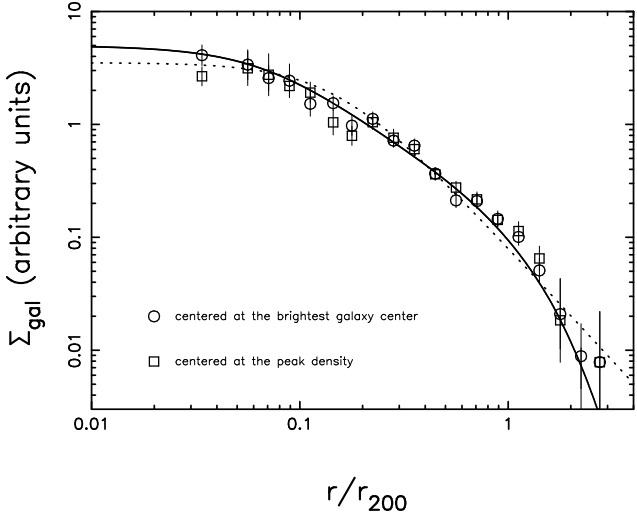,width=88mm,bbllx=86pt,bblly=86pt,bburx=564pt,bbury=642pt,clip=,angle=270}
     	\caption{Surface number density profile of galaxies over an 
 	ensemble of 14 distant clusters from CNOC (Carlberg, Yee, 
	\& Ellingson 1997). Solid line is our expected result of 
	the saturation model for $\sigma_d/\sigma_{gal}(0)=0.78$ and
 	$r_{200}/\sigma_d\tau=0.4$ under the assumption that galaxies 
	trace the gravitational potential of dark matter particles in 
	clusters. The conventional King model fit is illustrated 
	by the dotted line.
}
      \label{fig.4}
   \end{figure}

The magnitude of the predicted surface number density depends 
sensitively on the central velocity bias 
$\sigma_d/\sigma_{gal}(0)$. 
Its best fit value being less than unity may simply
reflect the fact that the central galaxy
velocity $\sigma_{gal}(0)$ is higher than the average value
$\overline{\sigma}_{gal}$.  To pin down the ratio of the averaged specific 
kinetic energy
of galaxies to that of dark matter particles for this value of
$\sigma_{gal}(0)$, we calculate
\begin{equation}
\frac{\sigma_d^2}{\overline{\sigma}_{gal}^2}=
	       \sigma_d^2\left/\left(
		\frac{\int_{x_b}^{\infty}
	       \sigma_{gal}^2(x)\Sigma_{gal}(x)xdx}
	       {\int_{x_b}^{\infty}\Sigma_{gal}(x)xdx}\right)\right.
\end{equation}

A straightforward calculation shows that 
$\sigma_d/\overline\sigma_{gal}\approx 1$.  That is, the best-fit central
velocity bias $\sigma_d/\sigma_{gal}(0)$ renders the average specific 
kinetic energies of both dark matter particles and galaxies to be approximately
equal, as required by our principle set forth to constrain the 
empirical parameters.  The good agreement is more than surprising for such a 
simple theoretical model.  Actually, it has
already been seen in the above subsection that the velocity
dispersion of dark matter particles ($\sigma_d=1300$ km s$^{-1}$),
required to account for the projected
mass distribution of clusters derived from gravitational lensing,
is roughly the same as the reported mean value for
the CNOC galaxies ($\overline{\sigma}_{gal}=1200$ km s$^{-1}$)).
This is also in agreement with the previous results given 
by a statistical 
comparison of cluster mass determinations from gravitational
lensing and velocity dispersion of cluster galaxies as the tracer
of cluster potential (Wu \& Fang 1997; Wu et al. 1998).

\subsection{Gas distribution and $\beta$ model}

Similarly, the equation of hydrostatic equilibrium can
also be applicable to the hot
X-ray emitting gas of number density $n_{gas}(r)$ in clusters
(Cavaliere \& Fusco-Femiano 1976):
\begin{equation}
\frac{1}{n_{gas}}\frac{d(\sigma_{gas}^2 n_{gas})}{dr}
 =-\frac{GM}{r^2},
\end{equation}
in which $\sigma_{gas}^2\equiv kT/\mu m_p$ is the ``equivalent
velocity dispersion'' of gas with $\mu m_p$ being the mean particle mass.
Assuming an equation of state for the gas
$T=T_0[n_{gas}/n_{gas}(0)]^{\gamma-1}$, we can write out the
gas number density profile in terms of eq.(14):
\begin{equation}
\frac{n_{gas}}{n_{gas}(0)}=
\left[1-\frac{\gamma-1}{\gamma}\int_{r_b}^{r}
\frac{GM}{\sigma_{gas}^2(0)r^2}dr\right]^{1/(\gamma-1)},
\end{equation}
where $r_b$ is the physical length corresponding to $x_b$.
Motivated by
the fact that $\overline{\sigma}_{gal}\approx\sigma_d$, the arbitrariness 
of the polytropic index $\gamma$ and $\sigma_{gas}(0)$ can be fixed by
a physical requirement that $\overline{\sigma}_{gas}$ also equals $\sigma_d$,
where $\overline{\sigma}_{gas}$
is the average equivalent velocity dispersion of gas:
\begin{equation}
\overline{\sigma}_{gas}^2=
		\frac{\int_{x_b}^{\infty}
	       \sigma_{gas}^2(x)S(x)xdx}
	       {\int_{x_b}^{\infty}S(x)xdx}.
\end{equation}
The observed X-ray surface brightness is
related to the temperature and number density of electrons in clusters as
\begin{equation}
S_x(r)\propto \int_{r}^{\infty}T^{1/2}(x)n_{gas}^2(x)
		     \frac{xdx}{\sqrt{x^2-r^2}},
\end{equation}
provided that the radiation mechanism is the thermal Bremsstrahlung
with a Gaunt factor weakly dependent on $T$.  

Before proceeding to a discussion on the general properties
of the theoretically predicted X-ray surface brightness profile,
we first present a comparison of the predicted profile with 
the observed X-ray surface brightness of an individual
cluster for an illustration of the agreement that also holds 
individually.  Figure 5 shows a typical example of
the {\it ROSAT} observed $S_x(r)$ for cluster Cl0016+16 ($z=0.55$),
together with our expected $S_x(x)$ from eqs.(15) and (17).
The latter are obtained by properly choosing the length scale parameter
$\sigma_d\tau$ and the central biasing parameter $\sigma_d/\sigma_{gas}(0)$
or equivalently the polytropic index $\gamma$.
This procedure is somewhat equivalent to the conventional $\beta$ model
fit by properly determining the X-ray core radius and the power index.
Yet, unlike the
$\beta$ model which is purely an empirical formula, our predicted
X-ray surface brightness profile is given by the hydrostatic
equilibrium of the intracluster gas with the underlying gravitational
potential of the cluster dark matter particles. 
The physics behind our prediction about the gas distribution
is well understood. 

    \begin{figure} 
	\psfig{figure=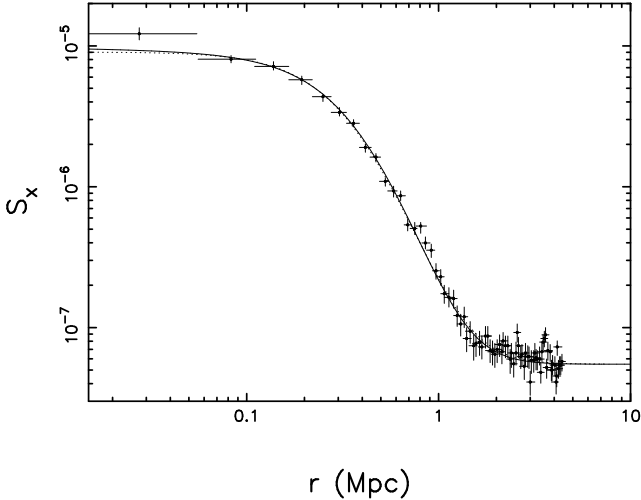,width=88mm,bbllx=86pt,bblly=86pt,bburx=564pt,bbury=642pt,clip=,angle=270}
	\caption{An examples of fit of the theoretically predicted
	X-ray surface brightness (solid lines) to the observation of
	Cl0016+16 (Neumann \& B\"ohringer 1997).
	The best $\beta$ model fit is also displayed (dotted lines) for
	comparison. A background surface brightness of
	$5.5\times10^{-8}$ s$^{-1}$ arcsec$^{-2}$ is assumed.
}
      \label{fig.5}
   \end{figure}

    \begin{figure*} 
      \psfig{figure=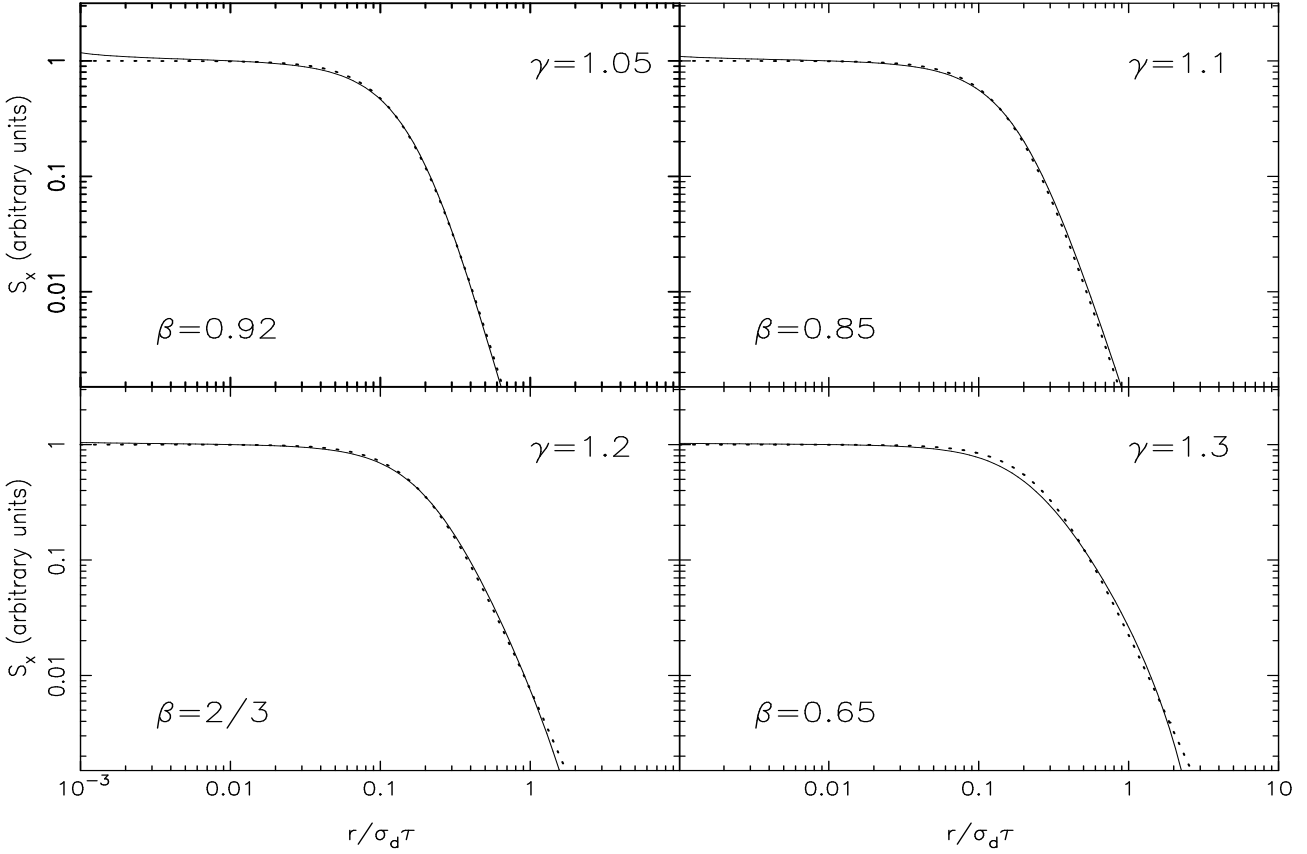,height=120mm,width=\textwidth,bbllx=50pt,bblly=50pt,bburx=580pt,bbury=736pt,clip=,angle=270}
	\caption{X-ray surface brightness profiles predicted for different
	polytropic indices $\gamma=1.05$ 1.1, 1.2 and 1.3 in the equation
	of state $T\propto n^{\gamma-1}$ (solid lines). The requirement that
	the mean kinetic energy of dark matter particles is equal to that of
	gas, $\sigma_d^2=(k\overline{T}/\mu m_p)$ has been employed.
	Also plotted are the
	conventional $\beta$ gas profiles (dotted lines).
}
      \label{fig.6}
   \end{figure*}

Fig.6 shows the X-ray surface brightness profiles of clusters obtained
from eqs.(15) and (17) for a set of polytropic indices $\gamma=$1.05,
1.1, 1.2 and 1.3, in which we have imposed a
constraint on $\sigma_d/\sigma_{gas}(0)$ by requiring
$\sigma_d/\overline\sigma_{gas}=1$, which uniquely defines a central
biasing parameter $\sigma_d/\sigma_{gas}(0)$ for each value of $\gamma$.
We now superpose the conventional
$\beta$ models on the theoretically expected $S_x$. Surprisingly, all the
derived $S_x$ profiles can be well approximated by
the $\beta$ models except near the compact cores, where brightness cusps
may appear in some cases.
The allowed $\beta$ parameters are dependent on the adopted
$\gamma$, with a smaller $\beta$ for a large $\gamma$.
Additionally, our fitted $\beta$ parameters are consistent with
the observed $\beta$ parameters in the range of
$\beta_{fit}\sim0.6$ -- $1.2$. If the average value of $\beta_{fit}$ is
indeed around $2/3$ as indicated by most current observations, we 
may conclude that a large proportion of galaxy clusters should have
a mean polytropic index of $\gamma\approx1.2$--$1.3$.
This point has recently been verified by the X-ray measurements of
temperature structures of 30 nearby clusters (Markevitch et al. 1998),
all of which show similar temperature profiles with a polytropic index of
$\gamma=1.2$--$1.3$.

On the other hand, the above fits for various $\gamma$'s yield that the core
radius of the gas profile varies from $r_c/\sigma_d\tau\sim0.16$
to $\sim0.30$.	Of these, the choice for
$\gamma=1.2$ that yields $r_c/\sigma_d\tau\approx0.20$
is again of particular interest. Since if the mean length scale $\sigma_d\tau$
is approximately $\sim1$ Mpc for rich clusters,
as was shown in the above two subsections,
this choice implies a mean core radius of $r_c\approx0.2$ Mpc in $S_x$; this
core radius is comparable to the median values (0.21--0.24 Mpc)
of the X-ray core radius distributions of the {\it ROSAT} detected
clusters (Vikhlinin et al. 1998).  Taking these results as a whole,
we have provided a consistent scenario for the physical origin of
the conventional $\beta$ model.

In sum, our fit suggests that the gas components of
clusters have an overall polytropic index close to $\gamma\approx1.2$
and a mean $\beta$ parameter $\overline{\beta}_{fit}\approx2/3$.  It is
worth stressing again that by requiring the ratio of mean specific 
kinetic energy in galaxy to that in gas to be unity,
$\beta_{spec}\equiv\overline{\sigma}_{gal}^2/(k\overline{T}_{gas}/\mu m_p)
 =\overline{\sigma}_{gal}^2/\overline{\sigma}_{gas}^2=
 (\sigma_d/\overline{\sigma}_{gas})^2/(\sigma_d/\overline{\sigma}_{gal})^2
 \approx1$,
the two $\beta$ values, $\overline{\beta}_{fit}$ and $\beta_{spec}$,
turn out to agree nicely with the observationally determined ones (Wu, Fang
\& Xu 1998).  The two $\beta$'s are not the same, 
and the so-called ``$\beta$ discrepancy''
(Bahcall \& Lubin 1994; references therein) appears to be a natural
consequence of our model.
We have also noticed that a similar
effort was recently made by Makino, Sasaki \& Suto (1998)
based on the universal dark matter profile.  However, while their derived
X-ray surface brightness is well represented by the $\beta$ model
in shape, their predicted core radius is a factor of (3--10) smaller
than the actually observed values.

\subsection{Baryon fraction}

Having known the dark matter and gas distributions, we can easily provide
the baryon (gas) fractions of clusters $f_b\equiv M_{gas}(r)/M_{total}(r)$,
where $M_{gas}$ and $M_{total}$ are the gas and total masses enclosed within
the radius $r$, respectively.
Of course, the absolute value of $f_b$ is ultimately related to
the amplitude of gas density profile, $n_{gas}(0)$, which
has remained unknown thus far.
Although we shall be able to figure
out $n_{gas}(0)$ using the observationally determined baryon fractions,
a particular attention is paid to the issue of how the baryon fractions
vary with scales.  Fig.7 demonstrates the cluster
baryon fractions versus radii obtained through the deprojection
analysis of {\it Einstein Observatory} data for 207 clusters
(White, Jones \& Forman 1997).
Basically, the baryon fraction is an increasing function with radius
and tends towards an asymptotic value of $\sim20\%$--$30\%$ at large
radius. Such a variation has been noticed previously by a number
of authors (e.g. White \& Fabian 1995; David 1997; White et al. 1997) but
not yet been satisfactorily accounted for
in the framework of any prevailing models for cluster formation.

Adopting a mean polytropic index of $\gamma=1.2$ for the X-ray gas, we attempt
to reproduce the observed radial variation of baryon fractions
with our model.  Since the length scale of our model has already been fixed,
the fitting can only be conducted in terms of the absolute value 
of baryon fractions.  The best-fit curve to is
illustrated in Fig.7.
The significant rise of $f_b$ with radius from $r\sim0.1$ Mpc
outward is due to the fact that
in this radial interval the dark matter profile goes as $r^{-3}$ (Fig.1)
whereas the gas distribution varies as $r^{-2}$ (Fig.6c). However,
it should be pointed out that the validity of our prediction about $f_b$
outside $r\sim5$ Mpc begins to be questionable,
though a decreasing tendency is seen to occur beyond $r\sim10$ Mpc.

Finally, we have noted that the majority of the data for the
baryon fractions $f_b$ have so far been derived from the isothermal
gas hypothesis, for which an emission-weighted temperature
over a finite X-ray emitting surface is applied.  Therefore, the comparison
between our prediction with the measured baryon fractions
holds true only statistically.

    \begin{figure} 
	\psfig{figure=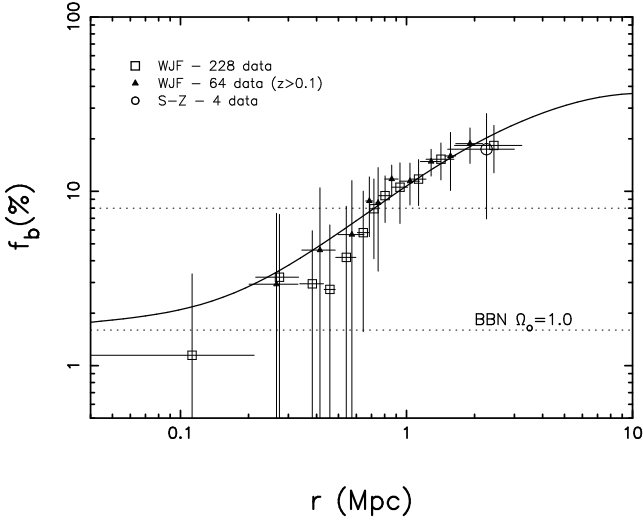,width=88mm,bbllx=86pt,bblly=86pt,bburx=560pt,bbury=642pt,clip=,angle=270}
	\caption{The binned radial variation of baryon (gas) fractions ($f_b$)
	of clusters. The data sets are from the {\it Einstein Observatory}
	(White, Jones, \& Forman 1997; WJF)	
	and from the measurements of the S-Z effects
	in four clusters (Myers et al. 1997). The high-redshift
	clusters ($z>0.1$) in WJF are explicitly plotted for comparison.
	The allowed  cosmological baryon fraction in the standard
	Big Bang Nucleosynthesis (BBN) for a matter dominant flat universe is
	also illustrated. The solid line is the theoretically predicted baryon
	fraction by our saturation model for gas with $\gamma=1.2$ and
	$\sigma_d/\sigma_{gas}(0)=0.62$
}
      \label{fig.7}
   \end{figure}

\section{Discussions and conclusions}

In the present work, we explore the possibility of building up two distinct
mass cores in a single galaxy cluster as a result of matter infall from the
background Hubble flow.   Comparisons of the predicted saturated
configuration with observations 
reveal that our model agrees quite well with the lensing, galaxy distribution
and gas distribution.
Our result is insensitive to the density parameter
$\Omega$ since the cluster builds up within a period much shorter than the age
of the universe.
The key assumptions of this model are (a) a
pre-existing mass concentration, onto which the background matter falls,
(b) the infall within the cluster is subsonic and the dark matter
particles are largely virialized with ergodic orbits, in which
situation the particle velocity dispersion is assumed to be a constant and (c)
in the saturated phase the mass ratio of the two cores remains the same as
that in the infall phase.

Despite of its oversimplification,
the self-similar gravitational collapse
of collisionless dark matter particles remains to be the only analytic approach
to access the dynamical process of cluster formation, which may
complement our understanding of its physical causes in addition to
the employment of large-scale numerical simulations.
In this paper we have re-addressed the self-similar infall model in a new
way: Our initial and boundary conditions are so chosen that there
is a pre-existent mass concentration in the central position of
cluster to induce mass accretion, and the process of major dark matter infall
is confined within a finite interval of $\sim10^{9}$ years.
Consequently, a hydrostatic equilibrium is established by now between the
gravitational potential of dark matter and the galaxies/gas in the
late evolutionary stage of clusters (probably since redshift $z\sim 0.8$).
In the scenario that the global properties of clusters
remain not evolving, we have successfully reproduced the
distributions of total mass, galaxies and hot gas in
clusters of galaxies revealed by gravitational lensing and optical/X-ray
observations.  Apart from the very central massive core of
$\sim$ few $10^{11}M_{\odot}$ within the radius of $\sim1$ kpc
(adopting $\sigma_d=10^3$ km s$^{-1}$ and $\sigma_d\tau=1$ Mpc), the
dark matter, galaxies
and gas essentially share a common ``secondary'' core at $r\sim0.15 - 0.2$ Mpc,
though their profiles can differ significantly, depending on
the radial variations of the velocity dispersion/thermal speed of
galaxies/gas.

We also find that the bias of the cluster averaged velocity dispersion of
galaxies against the velocity dispersion of dark matter 
is indeed approximately 
unity, i.e., $\overline{\sigma}_{gal}/\sigma_d\approx 1$, as expected 
from the virialization consideration that they are both within 
the same gravitational potential.
In view of this result, it is further demanded that
the specific dark matter kinetic energy and specific gas thermal energy to
also be equal, i.e., 
$(k\overline{T}_{gas}/\mu m_p)/\sigma_d^2\approx1$, thus yielding
the $\beta_{spec}$ parameter, which
denotes the ratio of specific kinetic energy
in galaxy to that in gas, to be also approximately unity
(Bahcall \& Lubin 1994; Girardi et al. 1996).  
The resulting X-ray surface brightness is found to be well fit 
by the conventional $\beta$-model, with,
however, the parameter $\beta_{spec}$ different from the
$\beta_{fit}$ parameter used in the $\beta$-model fitting of the X-ray surface
brightness.  The mean value of $\beta_{fit}$ turns out to be
$\overline{\beta}_{fit}\approx 2/3$, in contrast to $\beta_{spec}\approx 1$.
Therefore, the $\beta$ discrepancy, $\beta_{spec}/\beta_{fit} \ne 1$,
arises, in our model, as a natural consequence of the polytropic index
of gas $\gamma \ne 1$.  With the mean value
$\overline{\beta}_{fit}\approx 2/3$, it yields that the mean
value of $\gamma$ should be $\gamma\approx 1.2$.

Apparently, the present model has its own problems that remain to be resolved
in the future.  First of all, it is unclear at what epoch the
central massive core should begin to form; this issue is relevant to
the question as to when our model starts
to be applicable.  Second, although optical/X-ray observations
prefer a non-evolutionary scenario of galaxy clusters within $z\sim 0.8$,
we are unaware of how the background matters should terminate
the infall onto clusters.
It is possible that after the cluster mass reached some threshold value
the intercluster matters could then be heated up by the 
intracluster tidal forces exerted by clusters, thereby preventing further
matter infall. How this heating mechanism may proceed remains to be examined.
Finally, the massive compact core is predicted to be about $1$ kpc in size.
However, the current observations are unable to set any robust
constraints on the matter distribution inside the core region as small
as $r<10$ kpc.  It requires future observations
to confirm our prediction of compact cores at the cluster centers.

\begin{acknowledgements}
We gratefully acknowledge the insightful comments by an anonymous referee.
This work was supported by the National Science Council of Taiwan,
under Grant No. NSC87-2816-M008-010L and NSC87-2112-M008-009,
and the National Science Foundation of China, under Grant No. 19725311.
\end{acknowledgements}

\end{document}